\documentclass[aps,prd,preprint,longbibliography]{revtex4-2}

\usepackage{graphicx} 
\usepackage{soul}
\usepackage{xcolor}
\usepackage{plain}
\renewcommand{\hl}[1]{#1}

\long \def \blockcomment #1\endcomment{}

\begin{document}

\title{Chiral anomalies and Wilson fermions}
\author{Michael Creutz}
\affiliation{
  Senior Physicist Emeritus\\
  Physics Department\\
  Brookhaven National Laboratory\\
Upton, NY 11973, USA
}
\email{mike@latticeguy.net}

\date{\today}

\begin{abstract}
The Wilson formulation of fermions in lattice gauge theory provides a
unified description of the chiral anomalies in the standard model. The
discrete Dirac operator diagonalizes into a series of two by two
blocks.  In each block the possible eigenvalues either form a complex
pair or separate into two real eigenvalues that have specific
chirality.  The collision of these pairs of eigenvalues occurs outside
the perturbative region and provides a path between topological
sectors.
  \end{abstract}

\maketitle


\input epsf
\input colordvi  

\def \li {\par\hskip .2in \Maroon{$\bullet$}\hskip .1 in \textBlue}
\def \lli {\par\hskip .4in \Green{$\bullet$}\hskip .1 in \textMaroon}

\def \hi {\medskip \textBlack}
\def \topic #1{\par\textMagenta\centerline {#1}\ss\textBlack}

\def \nextslide{\bigskip}

\def\fitframe #1#2#3{\vbox{\hrule height#1pt
 \hbox{\vrule width#1pt\kern #2pt
 \vbox{\kern #2pt\hbox{#3}\kern #2pt}
 \kern #2pt\vrule width#1pt}
 \hrule height0pt depth#1pt}}

\long \def \blockcomment #1\endcomment{}
\def\slashchar#1{\setbox0=\hbox{$#1$}           
   \dimen0=\wd0                                 
   \setbox1=\hbox{/} \dimen1=\wd1               
   \ifdim\dimen0>\dimen1                        
      \rlap{\hbox to \dimen0{\hfil/\hfil}}      
      #1                                        
   \else                                        
      \rlap{\hbox to \dimen1{\hfil$#1$\hfil}}   
      /                                         
   \fi}                                         %


\section{Introduction}

The importance of chiral anomalies to the standard model has a long
history, going back to the calculation of the decay of the neutral
pion to two photons by Steinberger \cite{Steinberger:1949wx}.  Naive
arguments suggest that this should be suppressed when quark masses are
small.  This issue was elucidated in the classic papers of Adler,
Bell, and Jackiw \cite {Adler:1969gk,Adler:1969er,Bell:1969ts}.

In the context of the strong interactions, anomalies associated with
the axial current emerged as crucial to understanding why the eta
prime meson does not behave as a pseudo-Goldstone boson, unlike the
pions and kaons.  Also, a portion of the proton mass arises via a
similar mechanism.

Perhaps the most unexpected consequence of chiral anomalies is the
prediction by 't Hooft \cite{'tHooft:1976fv} that proton decay should
occur in the standard model of the weak interactions.  This rate is
extremely small, perhaps unmeasurable, but its existence is a
necessary consequence of the theory.

All these phenomena are tied to an anomaly with a simple change of
fermionic variables in path integral
\begin{equation}
  \matrix{ &\psi \longrightarrow
  e^{i\phi\gamma_5} \psi\cr &\overline \psi \longrightarrow
  \overline\psi e^{i\phi\gamma_5}\cr }.
  \label {axial}
\end{equation}
This is a symmetry of the classical standard model Lagrangian when
the fermion masses vanish.  However all renormalization schemes must
violate this symmetry.  In particular left
handed fields, $\psi_L=(1-\gamma_5)\psi/2$, ultimately must
mix with right handed ones, $\psi_R=(1+\gamma_5)\psi/2$, although
the mechanism for this depends on the details of the regulator.

The Wilson lattice approach provides a specific regularization making
the Feynman path integral mathematically well defined.  As such the
approach must allow for the effects of anomalies.  A summary of the
long history of this problem appears in Ref. \cite{Smit:2025tre}.  The
Smit-Swift model \cite{Smit:1985nu,Swift:1984dg} { (see also
  \cite{Smit:1986fn,Itoh:1986jh,Itoh:1987iy})} \hl{ is a particular
  attempt to formulate chiral gauge theories with Wilson fermions.
  When extended to include an entire generation of fermions, The
  approach provides} a specific lattice formulation for the standard
model \cite{Creutz:2023wxu}, based on three commuting gauge groups
$SU(3)\otimes SU(2).\otimes U(1)$.  Here we expand on that
presentation with more details on the connection between anomalies and
colliding eigenvalues of the Dirac matrix.  Among various interesting
issues is where does the baryon decay predicted by 't Hooft
\cite{'tHooft:1976fv} arise?  We will see that it comes from multiple
fermion combinations simultaneously coupling in a gauge invariant way
with complex configurations of Higgs and gauge fields.

A typical lattice gauge simulation generates configurations
for the gauge and Higgs fields and then explores Dirac operator on
these configurations.  In something like the hybrid Monte Carlo method
\cite{Duane:1987de}
for a dynamical simulation, feedback is included from the fermion
loops.  Here we will concentrate on the Dirac operator itself and not
be concerned with the details of generating the relevant gauge and
Higgs configurations.

\section {Doublers and the Wilson trick}

In lattice gauge theory, the free fermion action starts with the
derivative term replaced by nearest neighbor differences
\begin{equation}
 i\partial_\mu \gamma_\mu \psi  =
\longrightarrow 
\sum_{\mu} {i\gamma_\mu\over a}(\psi_{i+e_\mu}- \psi_i)
\end{equation}
The lattice sites are labeled $i$, space-time directions $\mu$, and
lattice spacing $a$.  Before including bosonic fields, this difference
takes a particularly simple form in momentum space:
\begin{equation}
p_\mu
\longrightarrow {1\over a}\sin(a p_\mu) =p_\mu+O(a)
\label{naive}
\end{equation}

The natural range of each component of lattice momentum is.
$-{\pi\over a} < p_\mu \le{\pi\over a} $ Unfortunately within this
range there is an extra pole in the propagator at $p_\mu \sim
{\pi\over a} $.  The Wilson \cite{Wilson:1974sk} solution to this
issue removes this pole by giving the doubler a large mass
\begin{equation}
i\gamma_\mu p_\mu \longrightarrow {1\over a}(i\gamma_\mu\sin(a p_\mu)
+1-\cos(a p_\mu))
\label{wilson}
\end{equation}
Effectively this adds a $O(ap^2)$ correction to the naive result, a
formally irrelevant operator $\sim \overline\psi\partial^2\psi$.

The added heavy states explicitly break chiral symmetry and are
ultimately the source of anomalies. One might think of them as an
analogy of the heavy states in a Pauli-Villars \cite{Pauli:1949zm}
approach.

\begin{figure}
 \centerline{ \includegraphics[width=.6\hsize]{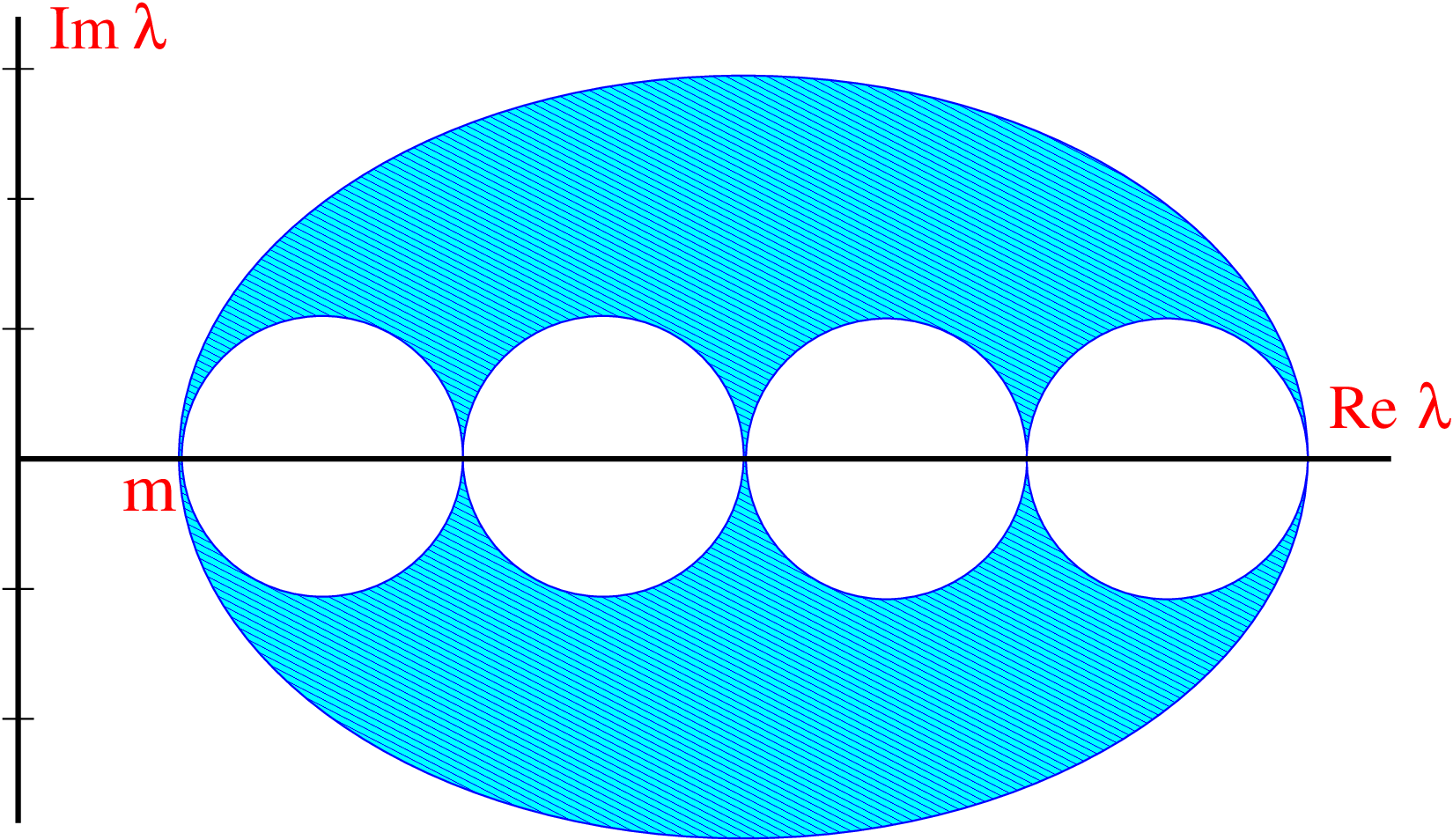}}
 \caption{ The spectrum of the free four dimensional Dirac operator.
   The crossings of the real axis correspond to how many doubler
   momentum components are at $\pi/a$.  The multiplicities at these
   points follow the pattern 1, 4, 6, 4, 1.  }
\label{grin}
\end{figure}

In the usual continuum theory, $i\slashchar D +m$ is an anti-hermitian
kinetic term plus a real constant mass term, As such, all eigenvalues
lie on a line at fixed real part $m$.  On the lattice the Wilson term
smears the mass term at high momentum.  For the free lattice theory
the sin and cosine factors in the above equations give rise to a set
of nested circles as shown in Fig. \ref{grin}.

When interactions turn on, the eigenvalues move around and momentum is
no longer a useful parameter.  Indeed, the Dirac operator is a
non-normal operator with $[D,D^\dagger]\ne 0$.  This comes about
because of the complex gauge factors as the fermions hop around.
Indeed, different eigenvectors of $D$ need not be orthogonal.  Also
the right and left eigenvectors are not generally the same.

\section{ Gamma-five hermiticity}

Although the lattice Dirac operator is neither Hermitian or normal,
it does satisfy what is called gamma-5 hermiticity
\begin{equation}
  \gamma_5 D= D^\dagger \gamma_5.
\end{equation}
The kinetic or ``hopping'' terms involving gauge fields are
anti-Hermitian and contain a gamma matrix factor while the mass and
Wilson term are Hermitian and commute with $\gamma_5$.

This ``symmetry'' has important consequences for the eigenvalue
spectrum of the Dirac operator.  Consider some particular complex
right {eigenvector $D\psi=\lambda\psi_1$}.  Playing with
the characteristic equation tells us
\begin{equation}
  0=|D-\lambda|=
|D^\dagger-\lambda^*|=
|D-\lambda^*|=
|D^\dagger-\lambda|
\end{equation}
Thus the complex conjugate of $\lambda$ is also an eigenvalue, and
both $D$ and $D^\dagger$ have the same eigenvalues.  Of course since
our matrix is not normal, the various eigenvectors are not in
general the same or even orthogonal.

To proceed it is convenient to consider a two dimensional space
spanned by our eigenvector {$\psi_1$ and $\psi_2\equiv \gamma_5
  \psi_1$}. Since $\gamma_5^2=1$, these each have the same norm, \hl
{which we can take to be unity.} By definition these satisfy
\begin{equation}
 \matrix{
 & D\psi_1=\lambda \psi_1\cr   
 & D^\dagger \psi_2=\lambda \psi_2.\cr   
} 
\end{equation}
 {We have a pair of right eigenvectors, one for $D$ and one for
  $D^\dagger$.}  Conjugation gives 
\begin{equation}\matrix{
 & \psi_1^\dagger D^\dagger=\lambda^*   \psi_1^\dagger\cr
 & \psi_2^\dagger D=\lambda^* \psi_2^\dagger.\cr
}
\end{equation}
{Thus we also have a pair of left eigenvectors, one for $D^\dagger$
  and one for $D$.  We see that $D$ maps pairs of right eigenvectors
  for $D$ and $D^\dagger$ into a similar pair of left eigenvectors.
  Furthermore, when $\lambda$ is complex, we have an orthogonality
  condition}
\begin{equation}
  \matrix{
&\psi_2^\dagger D \psi_1
=\lambda \psi_2^\dagger \psi_1
=\lambda^* \psi_2^\dagger \psi_1=0\cr
&\psi_1^\dagger D^\dagger \psi_2
=\lambda^* \psi_1^\dagger \psi_2
=\lambda \psi_1^\dagger \psi_2=0.\cr
  }.
  \label{conjugate}
\end{equation}
  The one combination we don't know is
  \begin{equation}
    \psi_1^\dagger D \psi_2 =(\psi_2^\dagger D^\dagger\psi_1)^*
    =\psi_1^\dagger D\gamma_5 \psi_1\equiv d
    \label{offdiag}
  \end{equation}
 where $d$ is an apriori unknown complex parameter.

{ The Dirac matrix thus breaks down into 2 by 2 blocks coupling
  pairs of right eigenvectors with pairs of left eigenvectors}
\begin{equation}
  \matrix{
    &\psi_i^\dagger D \psi_j=\psi_i^\dagger\pmatrix{\lambda & d \cr
                 0 & \lambda^*\cr}_{ij}\psi_j\cr
  &\psi_i^\dagger D^\dagger \psi_j=\psi_i^\dagger\pmatrix{\lambda^* & 0  \cr
      d^* & \lambda\cr}_{ij}\psi_j
    }.
\end{equation}
\hl{ These depend on two complex parameters, $\lambda$ and $d$.}

 This matrix has determinant $\lambda^* \lambda$ independent of $d$.
 This is a factor of the full determinant of $D$. Repeating this
 process on further eigenvalues, we find that the determinant of $D$
 can be expressed as a product of determinants of two by two complex
 matrices.
    
A complication occurs when the basis becomes singular with 
{$\psi_1\propto\psi_2$.  This can happen when the eigenvalues become real.}
To treat this situation it is useful to mix the modes into a
``chiral'' basis
\begin{equation}
  \matrix{
    &\psi_+= (\psi_1+\psi_2)/2
    ={1+\gamma_5\over 2} \psi_1\cr
    &\psi_-= (\psi_1-\psi_2)/2
    ={1-\gamma_5\over 2} \psi_1\cr
}.
\end{equation}
In this basis $\gamma_5$ reduces to the Pauli matrix $\sigma_3.$

We now write our sub-matrix of $D$ as the most general 2 by 2 matrix satisfying
$D^\dagger=\sigma_3 D \sigma_3$
\begin{equation}
D\rightarrow a_0+i a_1\sigma_1 +i a_2\sigma_2 +a_3\sigma_3
\end{equation}
where the coefficients $a_i$ are real.
This is easily diagonalized with eigenvalues
\begin{equation}
  \lambda=a_0\pm i\sqrt{a_1^2+a_2^2-a_3^2}
\end{equation}
The contribution to fermion determinant is the expected
\begin{equation}
  |D|=a_0^2+a_1^2+a_2^2-a_3^2=\lambda\lambda^*.
  \label{minkowski}
\end{equation}

The gamma-five hermeticity also implies
\begin{equation}
  H=\gamma_5 D
\end{equation}
is a hermitian quantity and always has real eigenvalues. This is also
contained in our 2 by 2 blocks, with eigenvalues
\begin{equation}
  \lambda_H=a_3\pm \sqrt {a_0^2+a_1^2+a_2^2}
\end{equation}
The real parts of the eigenvalues of $D$ are constrained to lie
between these.

\begin{figure}
 \centerline{ \includegraphics[width=.6\hsize]{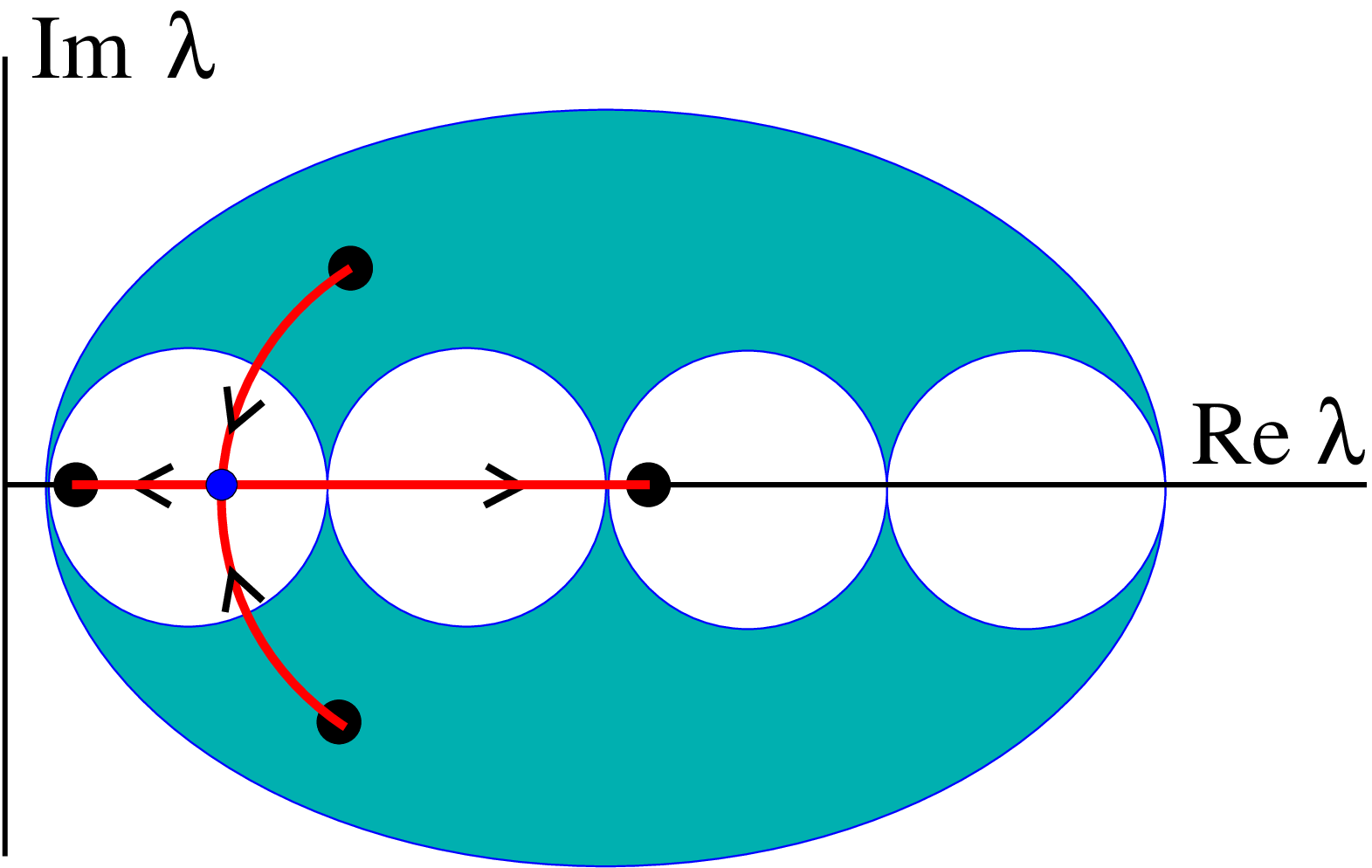}}
 \caption{ As gauge fields evolve, two complex eigenvalues can collide
   and move off on the real axis.  }
\label{collide}
\end{figure}

The parameter space of the four $a$'s separates nicely into two
domains.  When $a_1^2+a_2^2> a_3^2$ we have two complex eigenvalues as
above.  However when $a_1^2+a_2^2< a_3^2$ the eigenvalues become real
\footnote{Notice an analogy with the
connection between the Lorentz group and the group $SL(2)_C$, where
vectors also divide into two classes: time-like and space-like.}
\begin{equation}
  \lambda=a_0\pm
  \sqrt{a_3^2-a_1^2-a_2^2}.
\end{equation}
As the gauge and Higgs fields evolve in a simulation, we can transit
between these situations.  In the process the two eigenvalues collide
and move off in opposite directions on the real axis,as sketched in
Fig. \ref{collide}.

Large values of the parameter $a_3$ give rise to real modes of $D$.
If we now cool our configuration down to classical fields, these
become the topological objects commonly called ``instantons'' or
``sphalerons,'' the famous zero modes of the index theorem.  \hl{ With
  real eigenvalues, we lose the orthogonality condition of
  Eq. }\ref{conjugate}.\hl{ and the combination $\psi^* \gamma_5 \psi$
  need not vanish.  Indeed this must happen in the continuum limit for
  a single instanton.  In general real modes can mix right and left
  handed fermions.  This is at the heart of the anomalies that break
  chiral symmetry.  }

While real simulations do involve finite jumps in configurations, the
space of gauge and higgs fields is simply connected and there do exist
smooth paths between successive configurations.  Going between
different topologies must involve violating Luscher's "admissibility"
condition \cite{Luscher:2000zd}, as discussed in \cite{Creutz:2010ec}.
No assumptions are being made about the smoothness of the underlying
fields.

In our two dimensional sub-space, the role of $\gamma_5$ is played by
$\sigma_3$.  The free theory contains no terms directly involving
$\gamma_5$, so this term can only be generated from fermion loops in
the determinant of $D$.  Indeed a loop that involves hoppings in all
four dimensions can generate such a term.  Ref. \cite{Creutz:2010ec}
used the loops circumnavigating hyper-cubes to define a local
topological susceptibility.  See also Ref. \cite{Teper:1985rb}.  These
loops necessarily mix kinetic hoppings with hoppings involving the
Wilson term.

Note that the eigenvalue collision does not occur in perturbation
theory.  The effect does not occur until chiral effects are
sufficiently large.   Isolated modes with non-zero chirality
appear immediately after the real modes separate.  It is not required
to go to the zero mode limit.  Classical instantons and the index
theorem do prove chiral modes must exist {for smooth fields.
  Different lattice gauge configurations however are in general simply
  connected and these colliding eigenvalues provide a mechanism for
  flowing between different topologies.}

If one were to construct Neuberger's overlap operator
\cite{Neuberger:1997fp} by projecting all eigenvalues onto a circle,
singularities will appear as either of the modes pass through the
projection point \cite{Creutz:2002qa}.  In general these singularities
do not occur when the eigenvalues collide, but when either mode
crosses the circle center.

Small eigenvalues are formally suppressed in the path integral
\begin{equation}
Z=\int (dA)(d\overline \psi d\psi)
\ e^{-S_g+\overline\psi D \psi}
=\int (dA)\ e^{-S_g(A)}\ \prod \lambda_i
\end{equation}
Naively that would suggest that zero modes might be irrelevant.
However 't Hooft \cite{'tHooft:1976fv} showed how their effects
could be enhanced in physical observables.
This can be intuitively seen by adding source terms to the path integral
\begin{equation}
  \matrix{
Z(\eta,\overline\eta)&=&\int (dA)\ (d\overline \psi d\psi)\
e^{-S_g+\overline\psi D \psi +\overline\psi \eta+
\overline\eta\psi}\cr
&=&
\int (dA)\ e^{-S_g{+\overline\eta\ {D^{-1}}\ \eta}/4}\ 
\prod \lambda_i
  }
\end{equation}
The presence of $D^{-1}$ is how the zero modes remain relevant to
anomalies even as the quark masses go to zero.

\section{Physical consequences}

In a simulation $D$ is gauge gauge dependent; so, for physical results
we must combine the fields into gauge invariant observables.  As usual
in lattice gauge theory, we do not consider fixing a gauge.  As such
the various fields are highly fluctuating.  To avoid these
fluctuations we need to look at gauge invariant combinations of the
fields.

Here we concentrate in observables that are sensitive to the chiral
modes under discussion.  We briefly discuss effects with each of the
three commuting standard model groups $SU(3)\otimes SU(2)\otimes
U(1)$.

\subsection{Electrodynamics}

First we discuss electrodynamics $U(1)$ where it is well known that
one cannot conserve both the vector and axial current.  In this case
the above modes immediately break the axial symmetry of
Eq. \ref{axial}.  In this case it is perhaps simplest to think of the
doublers as playing the role of a Pauli-Villars field
\cite{Pauli:1949zm}.

\subsection{QCD}

Here we consider only one generation of fermions.  Thus we have $u$
and $d$ quarks which represent two SU(3) triplets.  When they are
degenerate the couple to exactly the same gauge and Higgs fields and
thus both see the same eigenvalues of $D$.  A chiral mode will flip
both spins together.  This generates an effective mass term for the
eta prime meson. 

\blockcomment
\begin{figure}
 \centerline{ \includegraphics[width=.6\hsize]{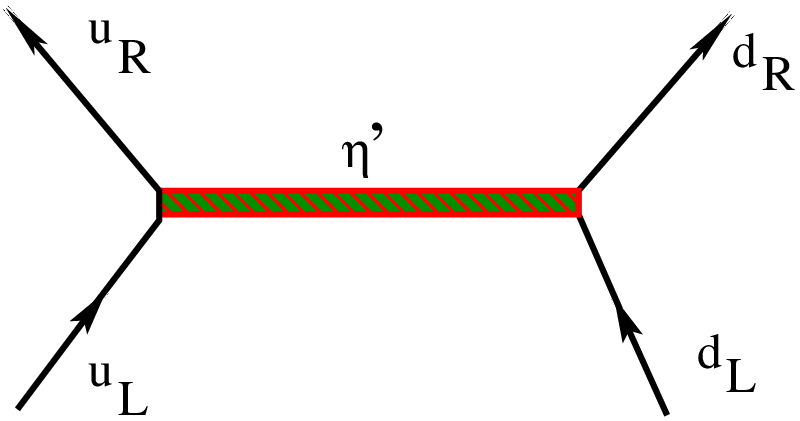}}
 \caption{As the up and down quarks flip spin from the chiral
   eigenmode, the eta prime acquires an effective mass.  {Without
     the anomalous eigenvalues in the path integral, the eta prime and
     neutral pion would be degenerate.  For the pion the effects
     cancel between the two flavors.}}
   \label{etaprime}
\end{figure}
\endcomment

The chiral modes can influence the masses of other gauge singlet
particles including the proton.  The effective operator affects each
quark color, generating an effective proton mass even if the quarks
are mass-less.

The magnitude of these effects is
controlled by the renormalization scale of the strong
interactions $\Lambda_{qcd}$
\begin{equation}m_{\eta^\prime},m_p\propto \Lambda_{qcd} \propto {1\over a}\
g^{-\beta_1/\beta_0^2}\ \exp\left({-1\over 2\beta_0 g^2}\right).
\end{equation}
This scale is not from classical instantons alone with
$\exp(-S_I)=\exp({-8\pi^2\over g^2})$, but rather is enhanced by
fermionic loops and quantum fluctuations.  The precise numberical
values can in principle be determined via numerical simulations.
 
\subsection {Weak $SU(2)$}

Now we turn to the group of the weak interactions.  Here 't Hooft
\cite{'tHooft:1976fv} has shown there is a small contribution to
baryon decay.  Because the basic electroweak coupling is small, the
effects are probably too small to observe.

The first generation standard model is built on four $SU(2)$ left
handed doublets
\begin{equation} \left(\matrix{u^r\cr
d^r\cr}\right)_L\ \left(\matrix{u^g\cr
    d^g\cr}\right)_L\ \left(\matrix{u^b\cr
    d^b\cr}\right)_L\ \left(\matrix{\nu\cr e^-\cr}\right)_L\
\end{equation}
where the superscripts $r,g,b$ refer to the strong interaction
``colors.''  Their antiparticles form four right handed doublets
$\psi^c_R=\tau_2\gamma_2\psi_L^*$.  Explicitly these are
\begin{equation}
\left(\matrix{d^{r,c}\cr r^{r,c}\cr}\right)_R\ 
\left(\matrix{d^{g,c}\cr r^{g,c}\cr}\right)_R\ 
\left(\matrix{d^{b,c}\cr r^{b,c}\cr}\right)_R\ 
\left(\matrix{e^+\cr \nu^c\cr}\right)_R\ 
\end{equation}
All of the above fermion variables have conjugate fields which are
independent Grassmann variables in the path integral.

When the gauge and Higgs fields of a configuration form the chiral
modes as discussed earlier, all of these doublets can be sensitive to
them.  To form gauge invariant observables, combine the left handed
doublets with the right handed ones to define a four by four matrix of
$SU(2)$ singlet combinations
\begin{equation}
  T_{ij}=\overline\psi_i^R\psi_j^L.
\end{equation}
To proceed we need to obtain a quantity that is also singlet
under the strong and electromagnetic groups.  This is accomplished in
the determinant of this matrix
\begin{equation}
  {\rm det} (T_{ij}).
\end{equation}
This antisymmetrizes over $SU(3)$ colors, making an $SU(3)$ singlet.
The quark and lepton charges also cancel making it electrically
neutral.  So this combination is is a fully gauge invariant operator
sensitive the chiral modes.  This is the determinant that 't Hooft
\cite{'tHooft:1976fv} found to summarize the weak anomalies in the
standard model.

This determinant includes terms that couple two quarks to an antiquark
and a lepton.  \footnote{This is consistent with the strong gauge
group since in $SU(3)$ the product of two triplet representations
contains an antitriplet, i.e. $3\otimes 3 = 6 \oplus \overline 3$.}
This allows for baryon-lepton mixing while preserving all gauged
symmetries.  This includes $p \leftrightarrow e^+$ and
$n\leftrightarrow \overline\nu$.

\section{Summary}

Given a configuration of gauge and Higgs fields in the path integral,
the corresponding Dirac operator can be reduced to { a series of 2
  by 2 matrices,} each representing two possible eigenvalues
contributing to the determinant of $D$.  As the configuration evolves
in a simulation, these modes can collide and transition between
complex pairs and real chiral modes.  This provides a common mechanism
for chiral anomalies in all three gauge symmetries of the standard
model.  These real modes occur outside the perturbative limit.
Cooling procedures to smooth fields drives such modes to known
topological configurations, although the anomalies do not require this
limit.

An interesting question concerns the counting of non-perturbative
topological parameters.  With Wilson fermions these are related to
relative phases between the fermion masses and the Wilson terms
\cite{Seiler:1981jf, Karsch:1985rc}.  In QCD flavored chiral rotations
can move this phase Theta between different quark species.  For
example it can be placed entirely in the phase of a single quark,
i.e. $m_u$ or even $m_t$.  How many such parameters are in the full
standard model?  Do both $SU(3)$ of QCD and $SU(2)$ of the weak
interactions have independent Theta parameters? And how does QED fit
into this picture?

\section*{Acknowledgment}
This manuscript has been authored under contract number
DE-AC02-98CH10886 with the U.S.~Department of Energy.  Accordingly,
the U.S. Government retains a non-exclusive, royalty-free license to
publish or reproduce the published form of this contribution, or allow
others to do so, for U.S.~Government purposes.


\begin{thebibliography}{10}

\bibitem{Steinberger:1949wx}
J.~Steinberger.
\newblock {On the Use of subtraction fields and the lifetimes of some types of
  meson decay}.
\newblock {\em Phys. Rev.}, 76:1180--1186, 1949.

\bibitem{Adler:1969gk}
Stephen~L. Adler.
\newblock {Axial vector vertex in spinor electrodynamics}.
\newblock {\em Phys.Rev.}, 177:2426--2438, 1969.

\bibitem{Adler:1969er}
Stephen~L. Adler and William~A. Bardeen.
\newblock {Absence of higher order corrections in the anomalous axial vector
  divergence equation}.
\newblock {\em Phys.Rev.}, 182:1517--1536, 1969.

\bibitem{Bell:1969ts}
J.S. Bell and R.~Jackiw.
\newblock {A PCAC puzzle: pi0 to gamma gamma in the sigma model}.
\newblock {\em Nuovo Cim.}, A60:47--61, 1969.

\bibitem{'tHooft:1976fv}
Gerard 't~Hooft.
\newblock {Computation of the quantum effects due to a four-dimensional
  pseudoparticle}.
\newblock {\em Phys.Rev.}, D14:3432--3450, 1976.

\bibitem{Smit:2025tre}
Jan Smit.
\newblock {A confederacy of anomalies}.
\newblock {\em Eur. Phys. J. H}, 50(1):5, 2025.

\bibitem{Smit:1985nu}
J.~Smit.
\newblock {Fermions on a Lattice}.
\newblock {\em Acta Phys. Polon. B}, 17:531, 1986.

\bibitem{Swift:1984dg}
P.~V.~D. Swift.
\newblock {The Electroweak Theory on the Lattice}.
\newblock {\em Phys. Lett. B}, 145:256--260, 1984.

\bibitem{Creutz:2023wxu}
Michael Creutz.
\newblock {Standard model and the lattice}.
\newblock {\em Phys. Rev. D}, 109(3):034514, 2024.

\bibitem{Duane:1987de}
S.~Duane, A.D. Kennedy, B.J. Pendleton, and D.~Roweth.
\newblock {Hybrid Monte Carlo}.
\newblock {\em Phys.Lett.}, B195:216--222, 1987.

\bibitem{Wilson:1974sk}
Kenneth~G. Wilson.
\newblock {Confinement of quarks}.
\newblock {\em Phys. Rev.}, D10:2445--2459, 1974.

\bibitem{Pauli:1949zm}
W.~Pauli and F.~Villars.
\newblock {On the Invariant regularization in relativistic quantum theory}.
\newblock {\em Rev. Mod. Phys.}, 21:434--444, 1949.

\bibitem{Note1}
Notice an analogy with the connection between the Lorentz group and the group
  $SL(2)_C$, where vectors also divide into two classes: time-like and
  space-like.

\bibitem{Creutz:2010ec}
Michael Creutz.
\newblock {Anomalies, gauge field topology, and the lattice}.
\newblock {\em Annals Phys.}, 326:911--925, 2011.

\bibitem{Teper:1985rb}
M.~Teper.
\newblock Instantons in the quantized su(2) vacuum: A lattice monte carlo
  investigation.
\newblock {\em Phys.Lett.}, B162:357, 1985.

\bibitem{Neuberger:1997fp}
Herbert Neuberger.
\newblock {Exactly massless quarks on the lattice}.
\newblock {\em Phys.Lett.}, B417:141--144, 1998.

\bibitem{Creutz:2002qa}
Michael Creutz.
\newblock {Transiting topological sectors with the overlap}.
\newblock {\em Nucl. Phys. Proc. Suppl.}, 119:837--839, 2003.

\bibitem{Seiler:1981jf}
E.~Seiler and I.O. Stamatescu.
\newblock {Lattice fermions and Theta vacua}.
\newblock {\em Phys.Rev.}, D25:2177, 1982.

\end{thebibliography}

\end{document}